\newcommand{\cns}{CE$\nu$NS}

\documentclass{ar-1col}

\usepackage{url}
\usepackage[numbers]{natbib}
\usepackage{rotating}
\usepackage{amsmath,amssymb}
\usepackage[letterpaper]{geometry}

\begin{document}

\markboth{Dutta and Strigari}{Neutrino physics with dark matter detectors}
\title{Neutrino physics with dark matter detectors}

\author{Bhaskar Dutta and Louis E. Strigari
\affil{Department of Physics and Astronomy, Mitchell Institute for Fundamental Physics and Astronomy, Texas A\&M University, College Station, TX, USA, 77843}
}

\begin{abstract}
Direct dark matter detection experiments will soon be sensitive to neutrinos from astrophysical sources, including the Sun, the atmosphere, and supernova. This sets an important benchmark for these experiments, and opens up a new window in neutrino physics and astrophysics. The detection of these neutrinos will be complementary to accelerator and reactor-based experiments which study neutrinos over the same energy range. Here we review the physics and astrophysics that can be extracted from the detection of these neutrinos, highlighting the potential for identifying new physics in the form of light mediators that arise from kinetic mixing and hidden sectors, and $\sim$ eV-scale sterile neutrinos. We discuss how the physics reach of these experiments will complement searches for new physics at the LHC and dedicated neutrino experiments. 
\end{abstract}

\begin{keywords}
Dark Matter, Neutrinos
\end{keywords}

\maketitle

\tableofcontents

\section{Introduction}
\par The detection of astrophysical neutrinos via new interaction channels, and in new energy regimes, has historically led to the discovery of physics beyond the Standard Model (SM), and to new insights into astrophysical sources. This has most famously been manifest in the fields of solar~\citep{Robertson:2012ib} and atmospheric~\citep{Fukuda:1998mi} neutrinos, which paved the way for the discovery of non-zero neutrino mass. The detection of neutrinos from SN 1987A~\cite{Hirata:1987hu,Bionta:1987qt} ushered in the era of extragalactic neutrino astronomy, and high energy, $\gtrsim$ TeV, neutrinos are now being identified from high redshift sources~\cite{IceCube:2018cha}.

\par Through a combination a terrestrial and astroparticle experiments, the mixing angles in the PMNS matrix that characterize neutrino flavor mixing are now well measured. Terrestrial experiments, as well as cosmological measurements of the large-scale structure in the universe, provide strong upper bounds on the sum of the neutrino masses, $\lesssim 1$ eV. However in spite of all the experimental progress, there are still questions that must be addressed in order to obtain a complete understanding of the physics of the neutrino sector. For example, is the neutrino mass hierarchy normal or inverted? What is the CP phase in the PMNS matrix? Does any data set, or combination of data sets, conclusively point to the existence of sterile neutrinos? These questions, along with the more general notion of searching for new physics through the neutrino sector, motivate the continued development of new classes of neutrino experiments. 

\par In the search for new physics that manifests itself in the neutrino sector, the low energy regime, characterized by momentum transfers less than $\sim$ MeV in the interactions between neutrinos and SM particles, presents an interesting path forward. At these energies, neutrinos have been detected from astrophysical sources, such as the Sun, the atmosphere, and supernova. Terrestrial sources, such as reactors and accelerators, are also used to study neutrinos at these energies. Though they have been studied for many years, neutrinos from reactors are still not completely understood, with some outstanding questions still lingering~\cite{Mention:2011rk}. Accelerator-based neutrino experiments also continue to pose interesting questions; for example there is not yet a conclusive theoretical explanation of the LSND excess~\cite{Aguilar:2001ty} that does not involve new physics. In reactor and accelerator-based experiments, neutrinos are specifically identified through elastic scattering interactions with electrons, and through inelastic interactions on nucleons and nuclei.  

\par These elastic and inelastic channels are not the only means through which neutrinos can be studied at $\sim$ MeV energies. As an example, a long-discussed, and theoretically well-studied, process is coherent elastic neutrino-nucleus scattering (CE$\nu$NS). The detection of CE$\nu$NS has very recently been achieved by the COHERENT experiment~\cite{Akimov:2017ade}, representing the culmination of many years of experimental efforts~\citep{Drukier:1983gj,Cabrera:1984rr}. It has opened up a new window into neutrino interactions in this energy regime, and along with it provides a new probe into beyond the SM physics. Interestingly, experimental proposals originally put forth to detect CE$\nu$NS were motivated to directly measure the different flavors of the solar neutrino flux. For SM interactions, the CE$\nu$NS process is equally sensitive to all neutrino flavors, making it an ideal detection channel through which the full solar neutrino flux could be studied. However before the detection of CE$\nu$NS, experiments such as SNO and Super-Kamiokande were able to measure all flavors of the solar neutrino flux, determining the combination of matter and vacuum-dominated oscillations to be the solution of the solar neutrino problem.  

\par Experiments constructed for the purpose of detecting CE$\nu$NS are closely related to those attempting to directly detect dark matter scatterings off of nuclei. Indeed, shortly after the experimental proposals were presented, it was determined that neutral current detectors sensitive to CE$\nu$NS can serve the additional purpose of detecting Weakly-Interacting Massive Particle (WIMP) dark matter~\cite{Goodman:1984dc}. Combining the cosmologically-motivated scattering cross section for WIMPs with nuclei with the estimation for the local number density of WIMPs, the first generation of low temperature germanium experiments were sensitive enough to detect WIMPs. While it was a distinct possibility that these experiments would detect WIMPs and render the non-gravitational detection of particle dark matter relatively straightforward, the first two experiments to systematically search for WIMPs reported null results~\citep{Ahlen:1987mn,Caldwell:1988su}. The constraints ruled out the simplest model for WIMP interactions with ordinary matter, and in the process first showed that if WIMPs comprise the dominant component of dark matter in galaxies, physics must be invoked to suppress the scattering cross section relative to that derived from the most basic cosmological arguments. In the subsequent several decades the limits on the WIMP-nucleus cross section has improved by nearly ten orders of magnitude~\cite{Akerib:2016vxi,Aprile:2018dbl}.  

\par Direct dark matter detection experiments are now coming full circle and are approaching the size at which they will be sensitive to astrophysical neutrinos from the Sun, atmosphere, and supernovae. While this will pose a new type of challenge for the detection of WIMP dark matter, it also presents new and exciting opportunities for studying neutrinos in a new realm. Extracting information about neutrinos from these forthcoming experiments requires a precise understanding of neutrino cross sections in this low $\sim$ MeV energy realm. It also requires detailing understanding of the sources and the fluxes of neutrinos that they produce. Identifying these neutrino interactions in future dark matter detectors paves the way for complementary studies with terrestrial experiments such as COHERENT, and also new reactor-based experiments that are attempting to detect \cns.   

\par In this article we discuss the opportunities for studying the physics of the neutrino sector, and the astrophysics of the neutrino sources, from the identification of neutrinos in future direct dark matter detection experiments. We stress that this new phase of experiments will be able to probe neutrinos in a way that has not yet been done with current neutrino experiments, highlighting a few examples of well-motivated beyond the SM models that these experiments will probe. 

\par The outline of this article is as follows. In Section 2, we review the basics of the neutrino cross sections for interactions of interest at these energies. In Section 3, we review the understanding of neutrinos from the Sun, supernova, and the atmosphere. In Section 4, we discuss the terrestrial experiments that have measured and are likely soon to measured CE$\nu$NS. In Section 5, we discuss the prospects for identifying new physics with the detection of neutrinos through CE$\nu$NS in dark matter and neutrino experiments. In Section 6, we discuss what can be expected in the upcoming years.  

\section{Low energy neutrino interactions}
\par In this section we review the neutrino interaction channels that will be most important for dark matter detectors: the CE$\nu$NS channel and the neutrino-electron elastic scattering channel. We review the SM predictions for these cross sections, and the simplest modifications to these cross sections that arise in beyond the SM physics in the form of non-standard neutrino interactions. In Section 5 below we discuss in more detail the possibility of beyond the SM contributions to these cross sections. 

\subsection{Coherent neutrino-nucleus scattering}
\par In the SM, neutrino-nucleon elastic scattering proceeds through the exchange of a $Z$ boson within a neutral current interaction. The resulting differential neutrino-nucleus cross section as a function of the nuclear recoil energy $T_R$ and the incoming neutrino energy $E_\nu$ is
\begin{equation}
\frac{d\sigma(E_\nu, T_R)}{dT_R} = \frac{G^2_f}{\pi} m_N  \left( Z g_v^p + N g_v^n \right)^2 
\left(1 - \frac{m_NT_R}{2E^2_{\nu}}
\right) F^2(T_R),
\end{equation}
where $m_N$ is the target nucleus mass, $G_f$ is the Fermi coupling constant, $N$ the number of neutrons, $Z$ the number of protons, $g_v^n = -1/2$, and $g_v^p = 1/2 - 2 \sin^2 \theta_w$, where $\theta_w$ the weak mixing angle. Note that the cross section is dominated by the neutron number. The nuclear form factor, which describes the loss of coherence due to the internal structure of the nucleus, is defined as $F(T_R)$. For momentum transfer less than the inverse of the size of the nucleus, the coherence condition is largely satisfied and $F(T_R) \rightarrow 1$. In lieu of experimental data on the neutron distribution in the nucleus, a typical parameterization for the nuclear form factor is the helm form factor~\cite{Lewin:1995rx}.

\par Due to the kinematics of the elastic scattering process, there is a predicted angular dependence of the recoil direction of the scattered nucleus. The directional and energy double differential cross section can be written by noting that the scattering has azimuthal symmetry about the incoming neutrino direction. Integrating over outgoing nuclear recoil energy gives
\begin{equation}
    \frac{d\sigma}{d\Omega} = \frac{G_f^2}{4\pi^2}
    \left( Z g_v^p + N g_v^n \right)^2
    E_\nu (1+\cos \theta) F^2(T_R) 
\end{equation}
where the angle is defined as $d\Omega = 2\pi \cos \theta d\theta$, where $\theta$ is the scattering angle between the incoming neutrino and the recoiling direction of the nucleus. 

\par The above expressions may be modified if there is physics beyond the SM that contributes to the scattering between neutrinos and quarks. Perhaps the simplest modification to the SM cross section comes from introducing non-standard neutrino interactions (NSI). Non-standard interactions may be motivated by introducing heavy new particles that mediate neutrino interactions with the SM. They are constrained by a wide range of neutrino experimental data; for a recent review of the constraints see Refs.~\cite{Ohlsson:2012kf,Miranda:2015dra}. For NSI,  the couplings, $\epsilon_{\alpha \beta}$, are defined as the effective strength of the new interaction relative to $G_f$, and the subscripts represent the new couplings for different neutrino flavors. For CE$\nu$NS, assuming neutral current vector-like interactions provides the modification to the cross section for an electron neutrino as
~\cite{Barranco:2005yy,Scholberg:2005qs}
\begin{eqnarray}
\frac{d\sigma(E_\nu, T_R)}{dT_R} &=& \frac{G^2_f}{\pi} m_N 
\left(1 - \frac{m_NT_R}{2E^2_{\nu}} \right)
F^2(T_R) 
\nonumber \\
&\times&
\left(
\left[ Z (g_v^p + 2 \epsilon_{ee}^{uV} + \epsilon_{ee}^{dV}) + N (g_v^n + 2 \epsilon_{ee}^{uV} + 2 \epsilon_{ee}^{dV})
\right]^2 
+ \sum_{\alpha = \mu,\tau} \left[Z (2 \epsilon_{\alpha e}^{uV} + \epsilon_{\alpha e}^{dV}) + N  (\epsilon_{\alpha e}^{uV} + 2 \epsilon_{\alpha e}^{dV})\right]^2
\right) \nonumber \\
\end{eqnarray}
Similar expressions may be derived for the $\mu$ and $\tau$ neutrino flavors. This equation implies that ~\cns~is sensitive to a combination of flavor diagonal terms and off-diagonal terms. Note that as defined the NSI couplings can be positive or negative. 

\subsection{Neutrino-electron elastic scattering}

\par The second important channel is neutrino-electron elastic scattering. In the SM, this process proceeds through both the exchange of a $Z$ boson and $W$ boson, where the latter is only possible in the case of an incoming $\nu_e$. The resulting cross section for a flavor $\alpha$ is
~\cite{Marciano:2003eq,Formaggio:2013kya} 

\begin{eqnarray}
   \frac{d\sigma_{\alpha}(E_\nu, T_e)}{dT_e} =  \frac{2 G_f^2 m_e}{\pi}
   \left[ (g_1^\alpha)^2 + (g_2^\alpha)^2 
   \left( 1 - \frac{T_e}{E_\nu} \right)^2 
   - g_1^\alpha g_2^\alpha \frac{m_e T_e}{E_\nu^2}
   \right]
\end{eqnarray}
where $m_e$ is the electron mass and $T_e$ is the recoil kinetic energy of the nucleus. For electron flavor, the couplings are defined as 
\begin{equation}
g_L^e = \sin^2\theta_w - \frac{1}{2} \ \ \ \ \ \ \ g_R^e =  \sin^2\theta_w
\end{equation}
For $\alpha = \mu, \tau$, the couplings are $g_1^{\mu \tau} = g_L^e$ and $g_2^{\mu \tau} = g_R^e$. Due to the relatively large difference in the $\nu_e + e$ and $\nu_{\mu,\tau}+e$ cross sections of almost an order of magnitude, by measuring the neutrino-electron scattering rate, for the case of solar neutrinos it is possible to derive the neutrino electron survival probability. 

\par Similar to the case of \cns~, NSI may impact neutrino-electron scattering, with the best sensitivity to non-universal flavor conserving couplings~\cite{Bolanos:2008km}. For the case of non-universal NSI, the couplings in the cross section are re-defined as 
\begin{equation} 
g_1^\alpha \rightarrow g_1^a + \epsilon_{\alpha \alpha}^{eL} \ \ \ \ \
g_2^\alpha \rightarrow g_2^a + \epsilon_{\alpha \alpha}^{eR}. 
\end{equation}

\section{Astrophysical sources}
\par In this section, we review the measurements and the theoretical interpretation of the neutrino fluxes from the Sun, the atmosphere, and from supernova. We highlight the progress that has been made in these areas in recent years, and how dark matter detection experiments may be used as probes of physics beyond the SM and of neutrino astrophysics.  

\subsection{Solar neutrinos}
\par
\begin{figure}[!htb]
\includegraphics[width=4.0in]{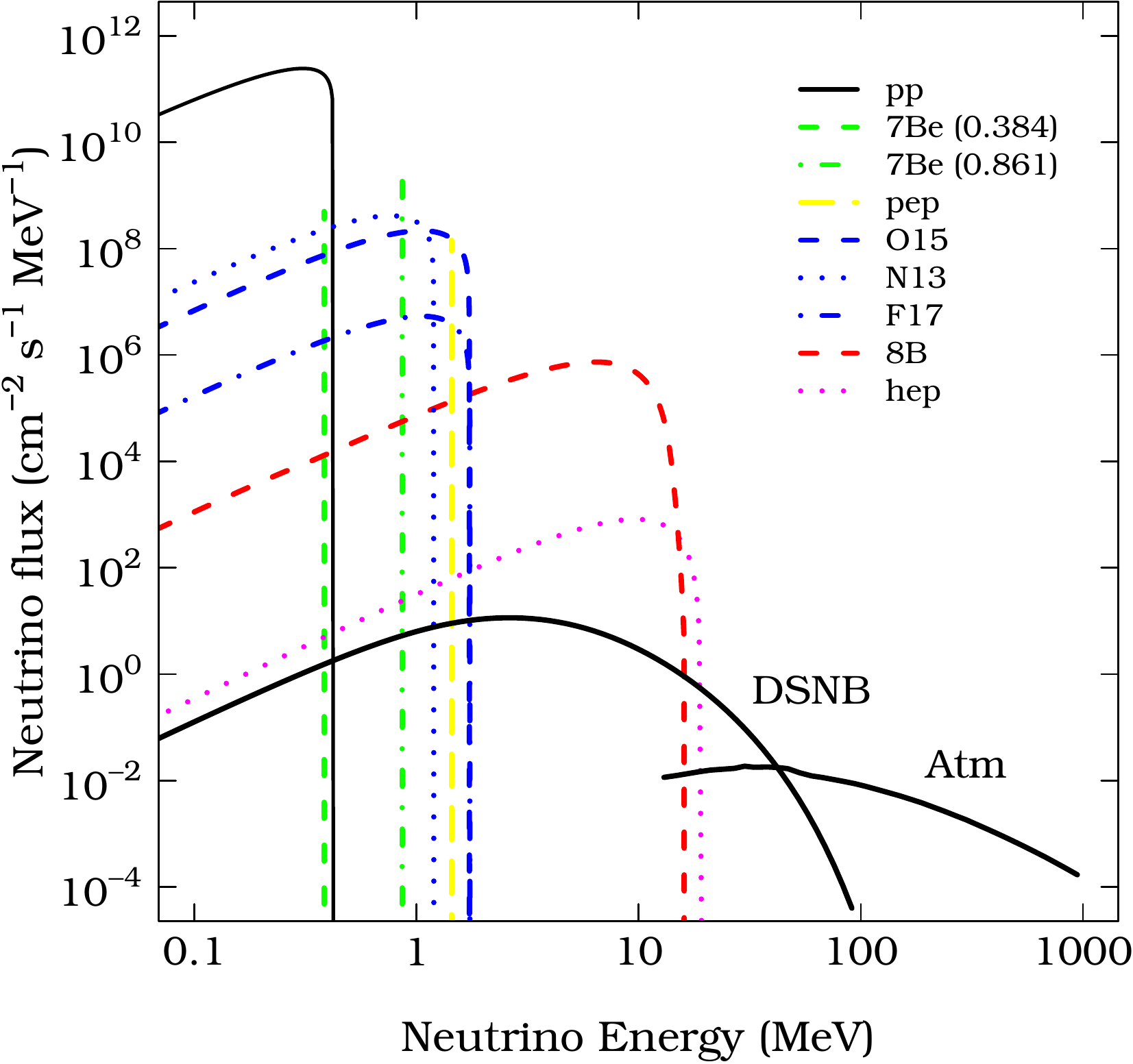}
\caption{Solar, atmospheric, and diffuse supernova neutrino fluxes. For the solar fluxes, each individual component is shown. For the atmospheric flux, shown is the sum of all four components ($\nu_e, \bar{\nu}_e, \nu_\mu, \bar{\nu}_\mu$). For the diffuse supernova neutrino fluxes, shown is the sum of all components ($\nu_e, \bar{\nu}_e, \nu_\mu, \bar{\nu}_\mu, \nu_\tau, \bar{\nu}_\tau$).} 
\label{fig:nuflux}
\end{figure}

 Experimental studies of solar neutrinos date back to over half of a century ago. The primary goal of these experiments is to measure the different components of the solar neutrino flux, as shown in Figure~\ref{fig:nuflux}, and to thereby provide an understanding of the physics of the solar interior. The original solar neutrino experiment, the Davis experiment at the Homestake mine in South Dakota, counted the number of electron neutrinos that captured onto Chlorine, $\nu_e + \textrm{Cl} \rightarrow e^- + \textrm{Ar}$~\cite{Cleveland:1998nv}. Due to the threshold for this capture process, this experiment was sensitive to the high energy $^{8}$B component of the solar neutrino flux. The measured counting rate was about one-third of that predicted by the Standard Solar Model (SSM) at the time. Subsequent Gallium experiments also detected electron neutrino captures, $\nu_e + \textrm{Ga} \rightarrow e^- + \textrm{Ge}$~\cite{Abdurashitov:2002nt}, and measured a counting rate of approximately one-half that predicted by the SSM. Due to the nature of this interaction Gallium experiments were sensitive to lower energy components of the solar neutrino flux; in addition to the $^{8}$B component they were also sensitive to the ${}^7$Be and pp components.

\par A significant advancement in the field of solar neutrinos came with the deployment of kilo-ton scale water cherenkov detectors. Experiments such as Super-Kamiokande~\cite{Abe:2010hy}, the Sudbury Neutrino Observatory (SNO)~\cite{Aharmim:2011vm}, and Borexino~\cite{Bellini:2011rx} measured the ``real-time" rate of neutrino-electron elastic scattering. Through neutrino-electron elastic scattering, these experiments were sensitive primarily to the electron neutrino flux, as well as some component of the muon and tau neutrino flux. With its measurement of the neutral current component of the solar neutrino flux through the break-up of deuterium, SNO determined that at high energies, nearly one-third of the $^{8}$B electron neutrinos transform into muon neutrinos in the Sun~\cite{Aharmim:2011vm}. Precision studies of the $^{8}$B component of the solar neutrino flux through elastic neutrino-electron scattering have continued, including a recent measurement of the $^{8}$B neutrino flux in the presence of ultra-low backgrounds by the SNO+ experiment~\cite{Anderson:2018ukb}. A global analysis of all solar neutrino data now finds $\lesssim 10\%$ uncertainty on the $^{8}$B component of the solar neutrino flux.

\par The Borexino experiment extended upon these results and has made precise measurements of the low-energy components of the solar neutrino flux. Borexino has now measured the rate of elastic neutrino-electron interactions from the 0.86 MeV ${}^7$Be line to $\lesssim$ 3\%~\cite{Bellini:2011rx,Bellini:2013lnn,Agostini:2017ixy}. The equivalent $\nu_e$-electron flux, which is calculated assuming that the observed rate is due only to electron neutrino interactions, is $(2.79 \pm 0.13) \times 10^9$ cm$^{-2}$ s$^{-1}$~\cite{Bellini:2013lnn}. Borexino has also measured the flux from the $p + e^- + p$ (pep) reaction, again from elastic neutrino-electron interactions~\cite{Bellini:2013lnn}. For pep neutrinos the equivalent $\nu_e$-electron flux is $(1.00 \pm 0.22) \times 10^8$ cm$^{-2}$ s$^{-1}$. Borexino has recently extended to even lower energy electron recoils, reporting the first direct measurement of the pp neutrino spectrum~\cite{Bellini:2014uqa}. This is the first spectral identification of the pp neutrino flux, and is consistent with the model that the Sun shines primarily from proton-proton fusion in the solar core. In addition to the ${}^7$Be, pep, and pp measurements, Borexino has placed an upper bound of $< 7.7 \times 10^8$ cm$^{-2}$ s$^{-1}$ on the sum of all components that contribute to the CNO flux.

\par In all, many different types of solar neutrino experiments have operated, and they have evolved in their size and scientific scope since the original Davis experiment. The combination of all solar neutrino data with terrestrial experiments that study neutrinos in the same energy range has led to the LMA-MSW solution to neutrino flavor transformation from the Sun to the Earth. With this solution, at low energies, $\lesssim 5$ MeV, vacuum oscillations describe the neutrino flavor transformation, and the electron neutrino survival probability is $\gtrsim 50\%$. At energies $\gtrsim 5$ MeV, matter-induced transformations describe the flavor transformation, with a corresponding survival probability of $\gtrsim 1/3$~\cite{Robertson:2012ib,Antonelli:2012qu}. 

\par However, in spite of all the theoretical and experimental progress in the field of solar neutrino physics over the past several decades, there are still some outstanding questions that surround some of the data. For example, three experiments (Super-Kamiokande, SNO, and Borexino) that are sensitive to electron recoils from neutrino-electron elastic scattering find that at electron recoil energies of a few MeV, the data are $\sim 2\sigma$ discrepant relative to the prediction of the best-fitting LMA-MSA solution. This may be indicative of new physics~\citep{deHolanda:2010am}, either in the form of sterile neutrinos or non-standard neutrino interactions. In addition, the recent measurement of the solar mass-squared difference from solar neutrino data, in particular from the day-night Super-Kamiokande data~\cite{Abe:2016nxk}, is discrepant at the $\sim 2 \sigma$ level relative to that measured by KamLAND~\cite{Gando:2010aa}. Non-standard interactions provide a possible solution to this discrepancy~\cite{Liao:2017awz}.

\par Another outstanding question relates to the measured neutrino flux, and how it is able to inform the physics of the solar interior. Recent theoretical modeling of solar absorption spectra and heliosiesmology data has suggested a lower abundance of metals in the solar core, i.e. a low-Z SSM~\cite{Asplund:2009fu}. This is in comparison to the previously-established high-Z SSM~\cite{Grevesse:1998bj}. Though some sets of solar neutrino data favor a high-Z SSM~\cite{Agostini:2017ixy}, a global analysis of all solar neutrino fluxes remains inconclusive~\cite{Bergstrom:2016cbh}.

\par Figure~\ref{fig:nuflux} shows the solar neutrino fluxes, with the normalizations of each flux corresponding to those from the high-Z SSM. The pp and pep are relatively insensitive to the assumed solar metallicity model, while the $^{8}$B, $^{7}$Be, and the CNO components are much more sensitive to the solar metallicity model. 
There is a particularly large theoretical uncertainty ($\sim 15\%$) on the CNO neutrino flux. 

\par
\begin{figure}[!htb]
\includegraphics[width=4.0in]{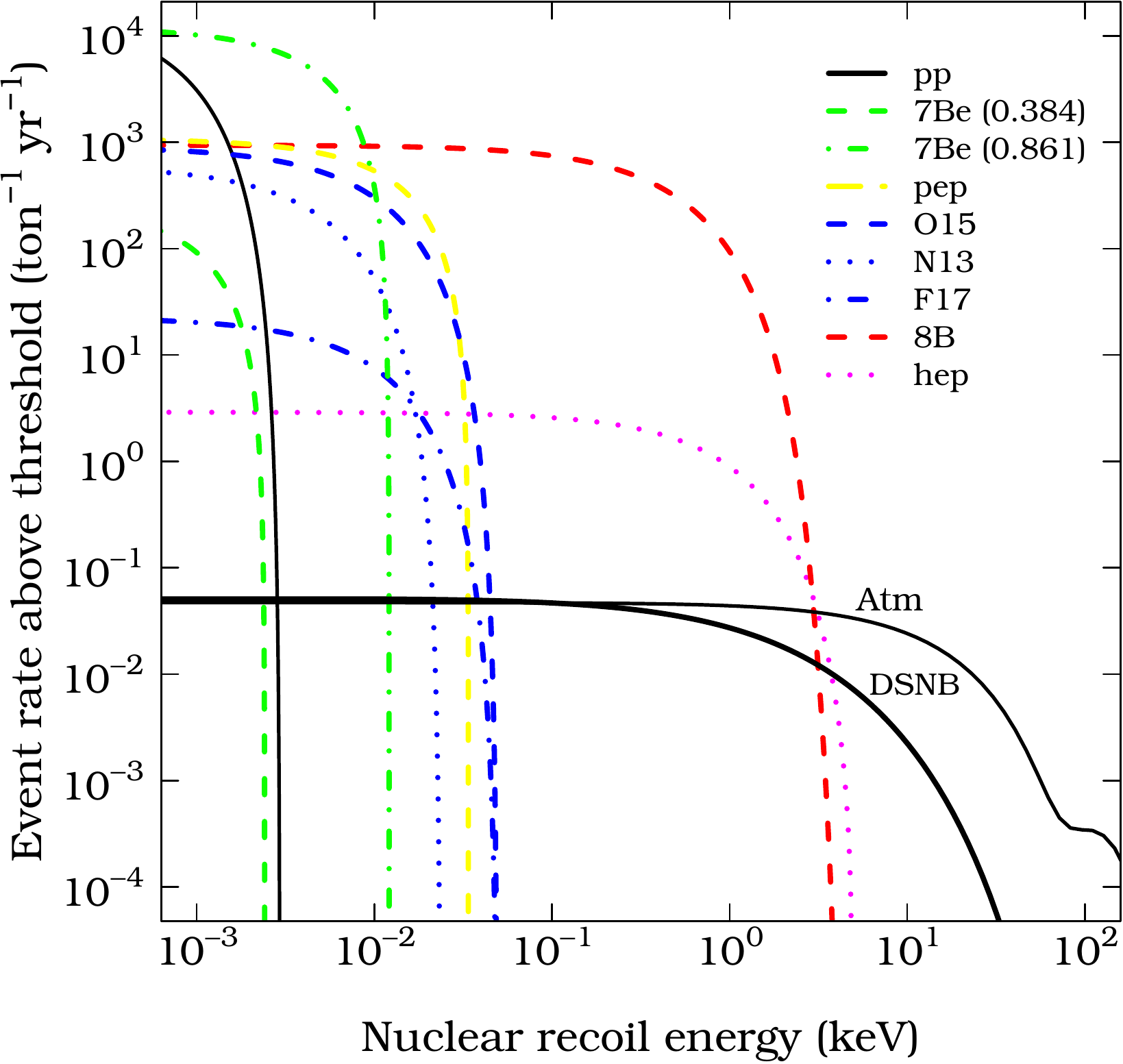}
\caption{Solar, atmospheric, and diffuse supernova neutrino event rates above a threshold recoil energy. For the solar fluxes, each individual component is shown. For the atmospheric flux, shown is the sum of all four components ($\nu_e, \bar{\nu}_e, \nu_\mu, \bar{\nu}_\mu$). For the diffuse supernova neutrino fluxes, shown is the sum of all components ($\nu_e, \bar{\nu}_e, \nu_\mu, \bar{\nu}_\mu, \nu_\tau, \bar{\nu}_\tau$).
} 
\label{fig:nuclearrates}
\end{figure}

\begin{figure}[!htb]
\includegraphics[width=4.0in]{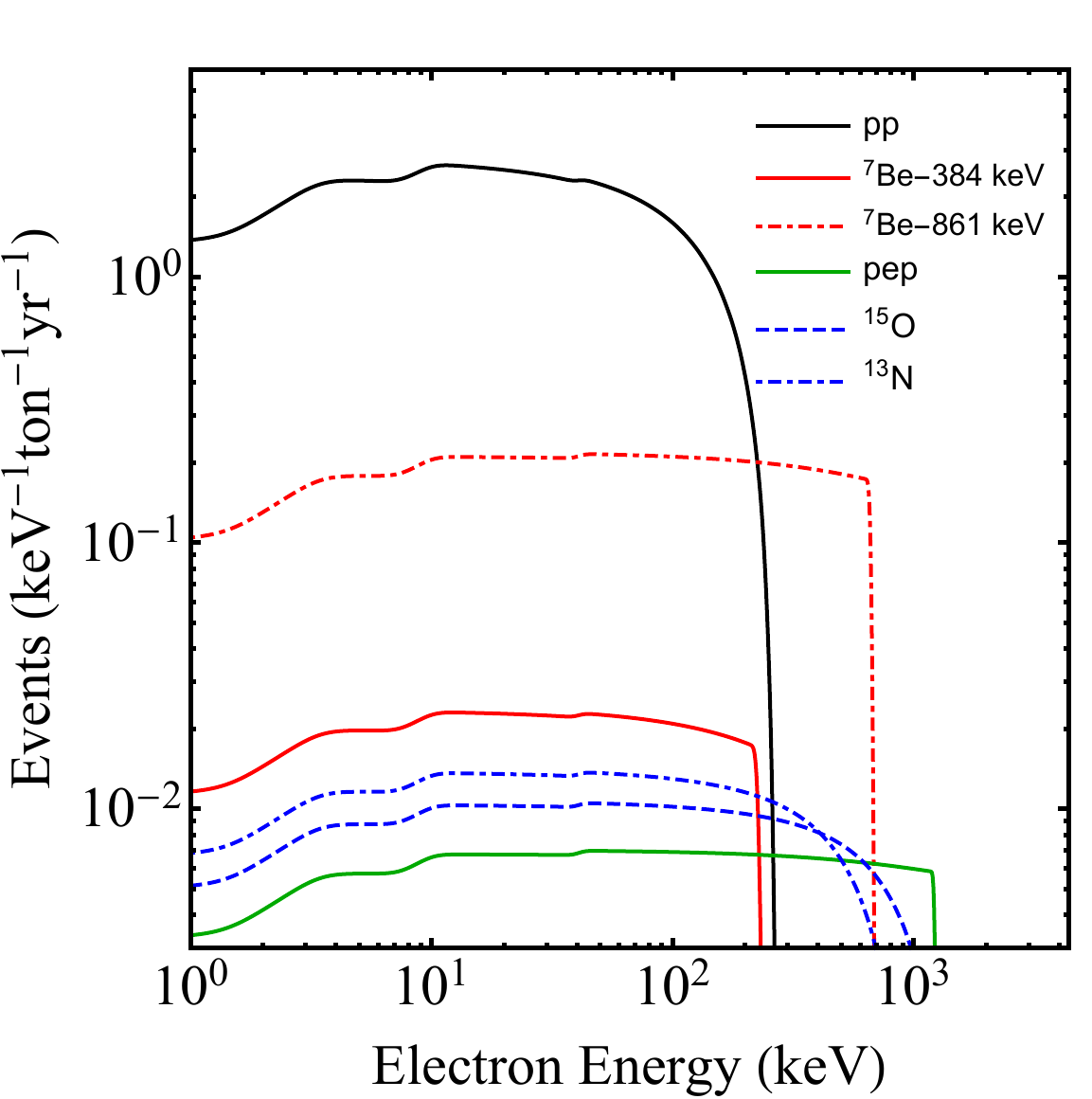}
\caption{Electron recoil rates from solar neutrinos for a Xe detector. Figure reproduced from Ref.~\cite{Newstead:2018muu}} 
\label{fig:electronrates}
\end{figure}

 From the fluxes in figure~\ref{fig:nuflux}, figure~\ref{fig:nuclearrates} shows the predicted nuclear recoil spectra, and figure~\ref{fig:electronrates} shows the predicted electron recoil spectrum for a xenon detector. For ton-year scale exposures, xenon-based detectors will have sensitivity to nuclear and electron recoils from the Sun. In the near future, the best chance of identifying a signal is probably the $^{8}$B solar neutrino-induced nuclear recoil events. As indicated, to measure the $^{8}$B component requires sensitivity to nuclear recoils $\lesssim 3$ keV.

\par Determining whether a xenon-based detector will be able to extract the nuclear recoil signal at these low energies requires an understanding of the detector characteristics. Liquid xenon 
detectors are dual phase, in that they detect both primary scintillation light and secondary light proportional to the amount of ionization from particles that scatter in the detector volume. A recoiling particle induces ionization and excitation of Xe atoms. For a given amount of energy deposited into the detector, recoiling electrons produce a much larger amount of ionization than recoiling nuclei. An electric field is employed in the detector that drifts the ionized electrons into a gas phase region, and there the electrons produce a signal proportional to the amount of ionization (this is the so-called ``$S_2$" signal). The primary scintillation signal (the so-called ``$S_1$" signal) results from the production of photons from the de-excitation of Xe atoms that have not been absorbed within the detector. Electron and nuclear recoils are then discriminated based upon the ratio of $S_2$/$S_1$. 

\par Nuclear recoil events with energies of $\lesssim$ 3 keV are typically detected with reduced efficiency relative to higher energy recoil events. This is because for a typical nuclear recoil at this energy, the primary scintillation S1 signal produces an amount of light that is below the threshold of the detector. So identifying events in this regime requires an upwards fluctuation in the amount of light yield that is greater than the mean yield for a nuclear recoil at this energy. Numerical software is now available that readily converts nuclear recoil energy into S1/S2 space for a given detector specification~\cite{Szydagis:2013sih}, thus allowing for a determination of the efficiency for extracting nuclear recoil events from solar neutrinos.  

\par Regarding electron recoils, the pp component is likely to be the first flux component that is measured. Characteristic electron recoil energies from this component are $\sim 1-100$ keV. For these higher energy electron recoils induced by solar neutrinos, there is a different set of experimental challenges. The most prominent background arises from the two-neutrino double beta decay (2$\nu \beta \beta$) of $^{136}$Xe. This is a rare decay process of $^{136}$Xe, which has a natural abundance in Xe of 8.9\%. The measured energy spectrum is the sum of the energy of the two outgoing electrons, with an endpoint of 2.459 MeV. The measured half-life for this decay is $\sim 2 \times 10^{21}$ years~\cite{Ackerman:2011gz}. The spectrum rises to high energies, such that there are $\sim 10^5$ events below 1.2 MeV. This 2$\nu \beta \beta$ background can be reduced through the use of xenon depleted of $^{136}$Xe. EXO~\cite{Albert:2017owj}, which aims to observe neutrinoless double beta decay, enriches the $^{136}$Xe of their target mass, and thus produces significant quantities of depleted xenon as a by-product. So it is plausible to this background by a factor of $\sim 100$~\cite{Newstead:2018muu}. This level of reduction is realized in the depleted xenon that is left over from $^{136}$Xe enrichment.

\par In addition to 2$\nu \beta \beta$, radioactive krypton represents another experimental background. Xenon that is extracted from the air is contaminated with trace amounts of $^{85}$Kr. Through cryogenic separation of the xenon~\cite{Aprile:2016xhi}, XENON1T has achieved the lowest levels of krypton concentration yet in their detector~\cite{Aprile:2018dbl}. For multi-ton scale next generation experiments, the krypton level may be even further reduced~\cite{Aalbers:2016jon}.

Yet another background to identifying electron recoils from solar neutrinos is $^{222}$Rn. This is a step in the uranium and thorium decay series, and has a half-life of 3.8 days. It continuously emanates from detector materials and readily mixes with the xenon target. Reduction of this has been achieved in the current XENON1T~\cite{Aprile:2018dbl} experiment, and for a next-generation detector this may be reduced even an order of magnitude further~\cite{Aalbers:2016jon}.

\par If the above experimental challenges can be overcome, direct dark matter detection experiments are poised to play an important role in the continuation of the solar neutrino program. In particular, they are in position to make the following measurements and scientific advancements in the forthcoming years: 

\begin{enumerate}
    \item An independent measurement of the neutral current component of the $^{8}$B solar neutrino flux. The only other existing ``pure" neutral current measurement of the $^{8}$B neutrinos comes from the SNO measurement of neutrino breakup of deuterium. The SNO neutral current data measured a flux that is right in between that predicted by the low and high Z SSM, and does not favor a particular model. Combining data from all experiments, the statistical uncertainty on the $^{8}$B flux is now about a factor of 5 less than the systematic uncertainty predicted by either SSM. Dark matter detection experiments will thus provide an important systematic check on the SNO result, and continue the program of using neutrinos to measure the conditions of the solar interior.  
    \item An independent measurement of the electron recoil energy spectrum induced by elastic scattering of pp neutrinos. While the pp neutrino flux was first indirectly identified as a component of the Gallium data, Borexino was the first experiment to make a measurement of the spectral energy distribution of electron recoil events induced by pp neutrinos. The Borexino measurement uncertainty on this component is now down to $\lesssim 10\%$. Further improving upon the measurement of this component will better constrain the "neutrino luminosity" of the Sun, as the pp neutrino flux is most directly related to the solar luminosity~\cite{Bahcall:2001pf}. This will also have the important effect of constraining alternative sources of energy production in the solar interior. With a reduction of electron recoil backgrounds at high energy described above, this measurement will be possible with future multi-ton xenon and argon-based experiments~\cite{Cerdeno:2016sfi,Newstead:2018muu}. 
    \item By extending to electronic recoils $\gtrsim 100$ keV, and for the case of a xenon detector depleting $^{136}$Xe, future multi-ton scale direct dark matter detection experiments may be able to make the first direct measurement of the CNO neutrino flux. This type of measurement has been discussed in the context of dedicated Ar detectors~\cite{Cerdeno:2017xxl}. 
    Measuring this flux component has been a holy grail of solar neutrino physics for many years, and can help to determine the fraction of the solar energy that is produced from the CNO cycle. The aforementioned Borexino limit on the CNO flux corresponds to a ratio of the flux with respect to the high-$Z$ SSM prediction of $< 1.5$. In absence of a planned, dedicated experiment to measure the CNO flux, dark matter experiments are one of the best bets to be the first to detect this component. 
    \item With a large enough exposure, future experiments may detect neutrinos from the minor branch of the pp chain the generates the most energetic neutrinos via the reaction $^{3}\textrm{He} + \textrm{p} \rightarrow ^{3}\textrm{He} + \textrm{e}^- + \nu_e$ (hep). Along with $^{8}$B neutrinos, neutrinos from the hep reaction also undergo adiabatic conversion in the solar interior. Neutrinos from the hep reaction have not been directly identified in solar neutrino experiments; the best upper bound from the SNO experiment is $\sim 4$ times greater than the SSM prediction~\cite{Aharmim:2006wq}. 
\end{enumerate}

\par The above discussion highlights the astrophysics that may be studied from a detection of solar neutrinos. The solar neutrino signal in direct dark matter detection experiments may also be used to search for the presence of new, beyond the SM physics. For example, a low-threshold ($\lesssim 1$ eV) ton-scale detector may be sensitive to an eV-mass scale sterile neutrino~\cite{Billard:2014yka}, which have been hinted at by several experiments~\cite{Mention:2011rk,Giunti:2010zu}. Future detectors will also be sensitive to new physics in the form of light mediators~\cite{Harnik:2012ni}, or enhanced baryonic currents~\cite{Pospelov:2012gm}. Xenon-based experiments currently have the sensitivity to probe non-standard neutrino interactions, which manifest themselves in the transition regime between vacuum and matter-dominated neutrino flavor transformations~\cite{Dutta:2017nht}.  These detectors are complementary to terrestrial searches for similar new physics scenarios, as described below. 

\par The discussion to this point has focused on detectors that are sensitive to nuclear and electronic recoils $\gtrsim 1$ keV. New classes of experiments are now under development that search for dark mater-electron interactions, which are able to detect sub-keV electron recoils that may be induced by dark matter $\lesssim$ GeV~\cite{Essig:2013lka}. In experiments which do not discriminate between nuclear and electron recoils, nuclear recoils from $^{8}$B will be the dominant background for these experiments~\cite{Essig:2018tss,Wyenberg:2018eyv}. In addition to $^{8}$B, Si-based experimental proposals will be sensitive to the $^{7}$Be component, providing the intriguing possibility of directly measuring the survival probability in the vacuum-dominated regime~\cite{Strigari:2016ztv}.

\subsection{Atmospheric neutrinos}
\par The collisions of cosmic rays in the atmosphere produce mesons and leptons, mostly in the form of pions and muons, across a wide range of energies. These heavy mesons and leptons then decay to ultimately produce a flux of muon and electron neutrinos and antineutrinos at the surface of the Earth. A precise determination of this atmospheric neutrino flux depends on several factors, including the cosmic-ray flux at the top of the Earth's atmosphere, the propagation of the cosmic rays through the atmosphere, and the decay of the mesons and muons as they propagate though the atmosphere to the Earth's surface. Since the flavors of neutrinos that are produced in the decays are known, theoretical models accurately predict the ratio of the flavor components of neutrinos across all energies. However, the normalizations of the fluxes differ depending upon the theoretical input.  

\par Dating back several decades since their initial detection~\cite{,Achar:1965ova,Reines:1965qk}, many experiments have estimated the flux of atmospheric neutrinos over nearly the entire energy range which they are produced. Most recently they have been studied in detail by experiments such as Super-Kamiokande~\cite{Richard:2015aua}, SNO~\cite{Aharmim:2009zm}, MINOS~\cite{Adamson:2012gt}, and Ice Cube~\cite{Aartsen:2013jza}. In these experiments, in the most common detection channel a neutrino interacts with a nucleus in or around the detector, creating a $\gtrsim$ MeV outgoing lepton (typically a muon) whose direction is reconstructed to tag a neutrino interaction. Through these types of detection, detailed measurements of atmospheric neutrinos have not only confirmed the basic prediction of neutrino production, but also have been important in identifying new physics. Most famously, nearly two decades ago Super-Kamiokande measured the ratio of muon to electron type events, and established vacuum-induced $\nu_\mu$ to $\nu_\tau$ transitions as the solution to the zenith angle dependence of this ratio~\cite{Fukuda:1998mi}. 

\par While the atmospheric neutrino flux for energies $\gtrsim 1$ GeV has been well studied by the aforementioned experiments, the low-energy flux of atmospheric neutrinos, $\lesssim 100$ MeV, is difficult to both theoretically model~\cite{Battistoni:2005pd} and to measure. Though the energy spectrum of neutrinos produced corresponds to that of muon and pion decay at rest, the absolute normalization of the flux is less well constrained due to the uncertainties that arise from several physical processes. For example, the cosmic ray flux at the top of the Earth's atmosphere differs from the cosmic ray flux in the interstellar medium. One reason for this is because of the solar wind which decelerates cosmic rays that enter into the heliosphere. A second reason is due to the geomagnetic field, which induces a cut-off in the low-energy cosmic ray spectrum. Detailed modeling of both of these effects implies that for energies $\lesssim 100$ MeV, the uncertainty on the predicted atmospheric neutrino flux is approximately 20\%~\cite{Honda:2011nf}. Due in particular to the cutoff in the rigidity of cosmic rays induced by the Earth's geomagnetic field at low energies, the atmospheric neutrino flux is larger for detectors that are nearer to the poles~\cite{Honda:2011nf}. 

\par In its search for diffuse supernova neutrinos, Super-Kamiokande has modeled the interaction of low-energy atmospheric neutrinos through two channels~\cite{Bays:2012wty}. The first component arises from the inelastic scattering of neutrinos on nuclei, which produces a continuum spectrum that rises with energy $\gtrsim 10$ MeV. These interactions result from neutrinos $\sim 100$ MeV. A second component comes from muon neutrinos and anti-neutrinos with energies $\sim 150$ MeV; these neutrinos create low-energy muons which decay to produce a michel spectrum of detected electrons. Dark matter detectors will be mostly sensitive to even lower energy neutrinos, $\sim 30-40$ MeV for a typical detector, so mapping the Super-Kamiokande measurements onto predictions for dark matter detectors still requires an extrapolation to even lower energies. 

\par The atmospheric neutrino flux is nearly isotropic, with small predicted deviations. Over all energies, the atmospheric neutrino flux peaks near the horizon, at zenith angle $\cos \theta \simeq 0$. At high energies, the flux is very nearly symmetric about $\cos \theta \simeq 0$, as at these energies the cosmic ray particles are more energetic than the rigidity cutoff. At low energies, the flux becomes asymmetric, as the flux of downward-going ($\cos \theta = 1$) neutrinos is lower than the flux of upward-going neutrinos ($\cos \theta = -1$). The FLUKA results provide an estimate for the angular dependence of the atmospheric neutrino rate~\cite{Battistoni:2002ew}. There is also a seasonal variation in the neutrino flux based on the atmospheric temperature which induces an additional time modulation. However the exact time dependence of this effect at different latitudes is not known and is likely too small to have a large observable effect. For this reason it is sensible to ignore both the angular and time dependence of the atmospheric neutrino flux and model it as isotropic and constant in time. 

\par The predicted nuclear recoil energy distribution from atmospheric neutrinos is shown in Figure~\ref{fig:nuclearrates}. Though as discussed above there is a mild angular dependence on the rate, when this flux is convolved with the angular dependence of the coherent neutrino-nucleus cross section, the angular dependence is washed out and the recoil spectrum depends only weakly on direction. The event rates indicate that a detector exposure of $\sim 20$ ton-yr will be required to begin to be sensitive to atmospheric neutrinos. Because the high energies of the nuclear recoils, the effect of the nuclear form factor becomes important; in particular variations from the standard helm form factor assumed to generate the rates in figure~\ref{fig:nuclearrates} have a significant impact on the predicted rate. Because the form factor is sensitive to the neutron distribution in the nucleus, and there are no laboratory measurements that have been made of this distribution in a nucleus like Xe or Ar, this form factor will likely remain a significant systematic uncertainty in determining the event rate. 

\subsection{Supernova neutrinos}
\par It has now been over thirty years since the detection of neutrinos from SN 1987A ushered in the era of extragalactic neutrino astronomy. The identification of nearly twenty neutrinos from this explosion confirmed the basic picture of core-collapse supernova explosions, in which over 99\% of the energy associated with the burst is carried away in the form of neutrinos of all flavors. Neutrinos are expected to emerge from the core of a supernova with a nearly Fermi-Dirac spectrum. In the limit of no neutrino flavor mixing, the temperatures for $\nu_e$, $\bar{\nu}_e$, and muon/tau flavors are expected to be $\sim 3,5,8$ MeV, respectively~\cite{Raffelt:2003en}.

\par The detection of supernova neutrinos has led to important constraints on the production of light particles, as well as the fundamental properties of neutrino themselves. Even though neutrino detectors have been operating nearly continuously in the three decades since SN 1987A, there has yet to be a detection from neutrinos from a Galactic supernova~\cite{Ikeda:2007sa}. The next supernova event in the Milky Way or in nearby galaxies is expected to provide unprecedented information on the physics of neutrino propagation from the SN core~\cite{Janka:2012wk}. For example, large water cherenkov detectors such as Super-Kamiokande will measure thousands of events, mostly through the charged-current inverse beta decay channel, and hundreds of events through various other elastic and inelastic channels~\cite{Scholberg:2012id}. 

\par Dark matter detectors will play an important role in supernova neutrino astrophysics. Through these detectors, all neutrino flavors will be studied through the CE$\nu$NS channel~\citep{Horowitz:2003cz,Lang:2016zhv}. For many years, this has been recognized to provide important information on the nature of stellar collapse~\citep{Freedman:1977xn}. By measuring the mean neutrino energies, tens of events are likely enough to constrain the explosion energy of the supernova, and to reconstruct the supernova lightcurve. This will be possible with currently-operating ton-scale detectors, and with next generation, multi-ton scale detectors the extracted physics will rival that of more traditional neutrino detectors~\cite{Lang:2016zhv}. 

\par In addition to the yield from a Galactic supernova event, potential 100-ton detectors may have sensitivity to the diffuse supernova neutrino background (DSNB)~\cite{Lunardini:2010ab,Beacom:2010kk}. Modern predictions are that the DSNB flux is approximately 6 cm$^{-2}$ s$^{-1}$~\cite{Horiuchi:2008jz}, including contributions from all neutrino flavors. In addition to being a probe on SN physics, the DSNB is an independent probe of the local core-collapse supernova and cosmic star formation rate~\cite{Hopkins:2006bw}. Though the DSNB has not been directly detected, there are strong upper bounds on the $\bar{\nu}_e$ component of the flux from Super-Kamiokande~\cite{Bays:2011si}. The best predictions for the flux of all flavors implies that dark matter detectors with exposures $\sim 100$ ton-year should be sensitive to the DSNB~\cite{Strigari:2009bq}. 

\subsection{Implications for direct dark matter detection}
\par Direct dark matter detection experiments will enter a new phase when they become sensitive to neutrinos from the aforementioned astrophysical sources. For a standard spin-independent dark matter-nucleon scattering cross section~\cite{Lewin:1995rx}, the nuclear recoil energy spectrum from $^{8}$B neutrinos closely matches that from a 6 GeV dark matter particle. Similarly, a $\sim$ 100 GeV particle matches the recoil spectrum induced by atmospheric neutrinos, and a $\sim 20$ GeV particle matches the spectrum of diffuse supernova neutrinos~\cite{Billard:2013qya}. A similar correspondence may be made for spin-dependent dark matter -nucleon interactions. Only for a Flourine target, which is sensitive to spin-dependent dark matter scattering through coupling to protons in the nucleus~\cite{Ruppin:2014bra}, is the neutrino spectrum different from the dark matter spectra over the entire dark matter mass range. 

\par
\begin{figure}[htbp]
\includegraphics[width=4.5in]{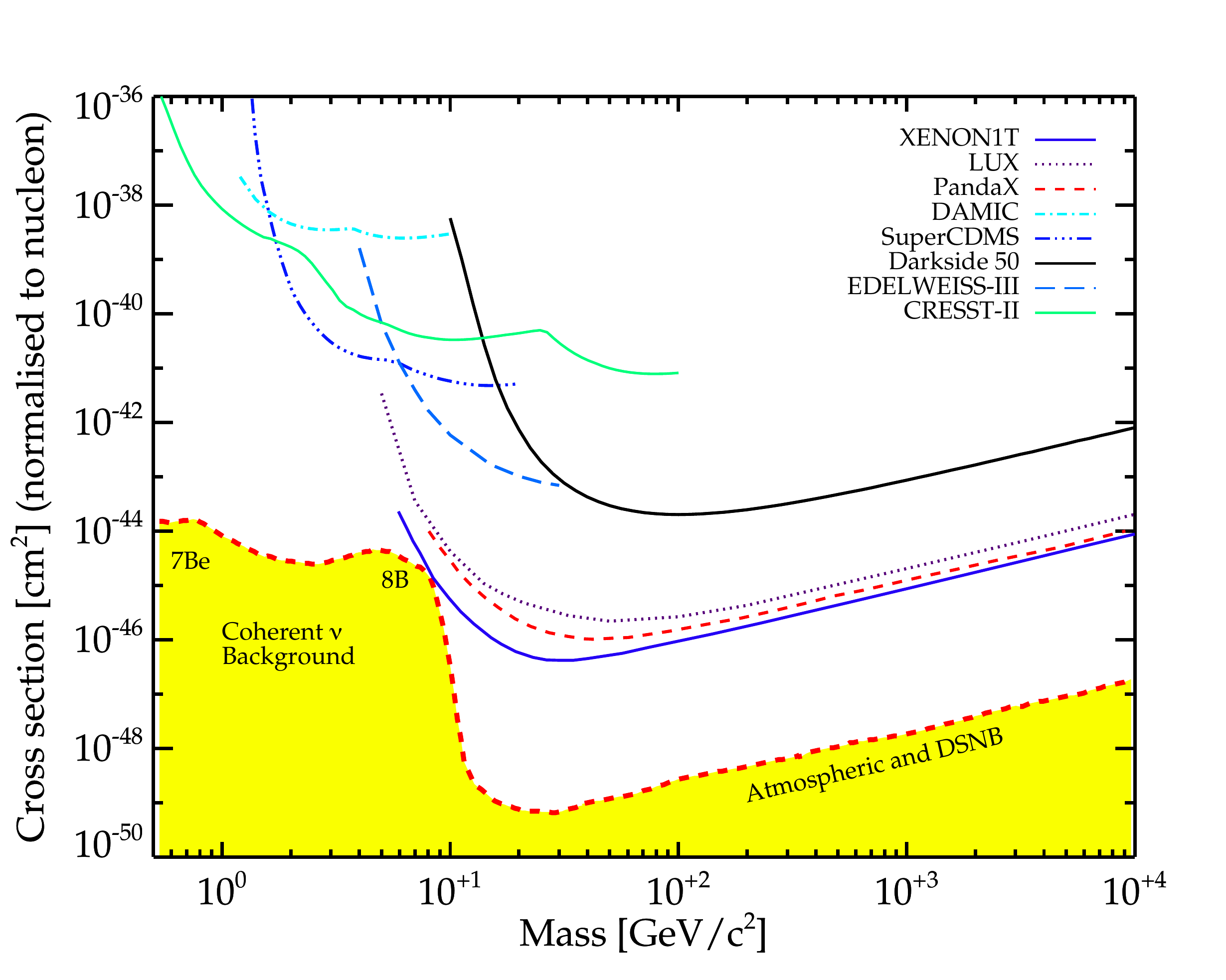}
\caption{Exclusion limits on the dark matter-nucleon spin-independent scattering cross section from various direct dark matter detection experiments. The shaded region shows the ``neutrino floor" as defined in Ref.~\cite{Billard:2013qya}}
\label{fig:nufloor}
\end{figure}

 Ultimately, neutrino-induced interactions will affect the detection of dark matter across the entire mass range of $\sim$ GeV to TeV. To get an idea as to the scale at which neutrino interactions will become important, using the event rate predictions~\cite{Monroe:2007xp,Vergados:2008jp,Strigari:2009bq}, Ref.~\cite{Billard:2013qya,Ruppin:2014bra} developed a statistical model to predict the corresponding value of the dark matter-nucleus scattering cross section, or equivalently the detector exposure, at which the sensitivity to the cross section saturates due to the neutrino backgrounds. This procedure defines a ``neutrino floor," which is shown as the shaded region in Figure~\ref{fig:nufloor}. 

\par At the highest nuclear recoil energies, $\gtrsim 20$ keV, the neutrino floor is determined mostly by low-energy atmospheric neutrinos described above. These will limit the sensitivity of dark matter detectors without directional sensitivity to spin independent cross sections greater than approximately $10^{-48}$ cm$^2$~\cite{Strigari:2009bq,Billard:2013qya,Ruppin:2014bra}. At the lowest nuclear recoils, $\sim$ a few keV, the floor is set by $^{8}$B solar neutrinos. These will limit the sensitivity of dark matter detectors without directional sensitivity to spin independent cross sections $\gtrsim 10^{-44}$ cm$^2$. As is seen, several experiments are now approaching this sensitivity, especially in the $^{8}$B solar neutrino energy regime. 

\par It is clear that in order to continue the development of the direct dark matter detection program, experiments must be able to identify neutrino-induced recoils, and to distinguish them from dark matter-induced events. Several methods have been proposed to distinguish neutrinos from dark matter scatterings in future detectors: 
\begin{enumerate}
    \item For the particular case of solar neutrinos, it is conceptually the simplest to employ detectors with sensitivity to the direction of the nuclear recoil. Studies have shown that for a $\sim$ ton-scale detector, neutrinos can be cleanly distinguished from dark matter~\cite{Grothaus:2014hja,OHare:2015utx}. The issue at this stage is one of feasibility; detectors that can measure the direction of nuclear recoils have sensitivity orders of magnitude weaker than those that are just sensitive to energy deposition alone~\cite{Mayet:2016zxu}. Similarly, a polarized detector would be able to discriminate dark matter from solar neutrino-induced events~\cite{Franarin:2016ppr}.
    \item For a standard spherically-symmetric and isotropic dark matter halo, the dark matter-induced nuclear recoil spectrum is expected to modulate annually~\cite{Freese:1987wu,Freese:2012xd}. The dark matter signal peaks in June and is at a minimum six months later. For comparison, due to the eccentricity of the Earth's orbit, the solar neutrino signal peaks in January, and is at a minimum six months later. The dark matter modulation effect is at the $\lesssim 10$\% level, while the neutrino modulation effect is at the $\lesssim 5\%$ level. Multi-ton scale detectors with good energy resolution may be able to discriminate between dark matter and neutrino-induced recoils using these modulation signals~\cite{Davis:2014ama,OHare:2015utx}. 
    \item For standard assumptions for the nature of the spin-independent and spin-dependent dark matter-nucleon interactions, the precise differences between the dark matter and neutrino-induced recoil spectra may be exploited, if a large enough detector is constructed. Refs.~\cite{Strigari:2009bq,Ruppin:2014bra} find that the appropriate detector exposure is nearly five orders of magnitude larger than the $\sim$ ton-yr exposure scales that are currently employed. 
    \item The theoretical framework that underlines dark matter-nucleon interactions may be more rich than the standard spin-independent/spin-dependent interaction formalism that experiments typically assume to derive limits on the cross section. For example, according to the non-relativistic effective field theory (EFT) framework~\cite{Fan:2010gt,Fitzpatrick:2012ix}, a combination of up to 30 dark matter nucleon operators may describe their interactions. For many of these operators the nuclear recoil spectrum from a dark matter scattering is distinct from the solar and atmospheric neutrino-induced spectrum. Further, the operators may be grouped based on their predicted nuclear recoil spectra~\cite{Dent:2016wor,Dent:2016iht}. Thus future detectors with precise nuclear recoil energy resolution can distinguish between nuclear recoils induced by dark matter and neutrino interactions. These interactions may also have unique signatures in searches with directional sensitivity~\cite{Catena:2015vpa}. 
\end{enumerate}

\par New physics, in particular in the form of NSI, also affects the discovery limit for dark matter in light of the neutrino backgrounds~\cite{Boehm:2018sux,Gonzalez-Garcia:2018dep}. In its simplest implementation, NSI induce an uncertainty on the normalization of the neutrino event rate, and an increased uncertainty on the rate has the effect of reaching the background saturation point at a larger cross section~\cite{Ruppin:2014bra}. The discovery limit for inelastic dark matter is also predicted to be different than for the case of standard spin-independent/spin-dependent dark matter~\cite{Gelmini:2018ogy}. Similarly, uncertainties on the astrophysical parameters, such as the local dark matter density and the circular velocity, also induce a systematic uncertainty in the location of the neutrino floor~\cite{OHare:2016pjy}.

\section{Terrestrial experiments}
The identification of neutrinos in dark matter experiments, in particular through the CE$\nu$NS channel, will be complementary to terrestrial neutrino experiments which operate in a similar energy regime. In this section, we review the various terrestrial CE$\nu$NS detection methods, focusing in particular on accelerator and reactor sources, which have been traditionally used to study neutrino properties in the MeV energy regime. We highlight the implications of the very recent CE$\nu$NS detection by the COHERENT experiment, as well as advances toward detection by the various reactor-based experiments. 

\subsection{Accelerator-based experiments: COHERENT}
\par The COHERENT experiment uses a stopped-pion source of neutrinos generated by the Spallation Neutron Source (SNS) at the Oak Ridge National Laboratory~\cite{Akimov:2018ghi}. Muon neutrinos with energy 30 MeV are produced from the charged pion decays, and $\bar{\nu}_\mu$ and $\nu_e$ are produced with a Michel energy spectrum from the subsequent decay of muons at rest. Due to the decay lifetime, $\bar{\nu}_\mu$ and $\nu_e$ from muon decays are delayed relative to the 30 MeV $\nu_\mu$ neutrinos produced from the prompt pion decay. With characteristic energies of tens of MeV, the coherence condition for the momentum transfer to the nucleus is expected to be preserved for a large enough sample of the neutrino-nucleus interactions~\cite{Kerman:2016jqp,Bednyakov:2018mjd}. The neutrino source at SNS is pulsed, which aids in the identification of backgrounds. 

\par In the summer of 2017, using 14.6-kg CsI[Na] scintillator detectors, the COHERENT collaboration announced the first detection of CE$\nu$NS~\cite{Akimov:2017ade}. COHERENT measured a best-fit count of 134$\pm$22 CE$\nu$NS events, well in excess of the expected backgrounds. The measured rate was found to be 77$\pm$16 percent of the SM prediction.

\par From this initial detection, COHERENT was able to constrain NSI in a regime of parameter space that had not been possible to probe. In particular the COHERENT data is sensitive to u and d-type NSI for flavor-diagonal muon components, $\epsilon_{\mu \mu}$. The COHERENT detection also sets new constraints on exotic solutions to solar neutrino mixing~\cite{Coloma:2017egw,Coloma:2017ncl}, and the g-2 anomaly~\cite{Liao:2017uzy}. The detection places novel constraints on new physics that manifests through the neutrino sector, including generalized scalar/vector neutrino interactions~\cite{AristizabalSierra:2018eqm}. It is also possible to constrain the neutron form factor for CsI~\cite{Ciuffoli:2018qem}, and sterile neutrinos~\cite{Kosmas:2017zbh}. Additional implications for new physics will be discussed below in Section 5. 

\subsection{Reactors}
\par Dating back to the first detection of neutrinos, nuclear reactors have been purposed as a copious source of electron anti-neutrinos. The characteristic neutrino energy is $\lesssim 1$ MeV, which is nearly an order of magnitude less than the neutrinos produced by accelerator sources. Due to these low energies, the coherence condition for the recoil is largely preserved over the entire reactor energy regime, so that there is no dependence on the internal structure of the nucleus~\cite{Bednyakov:2018mjd}. To this point, the primary difficulty in detecting \cns~using reactors is that detectors have not been able to achieve the low threshold required to identify the nuclear recoil signal. 

\par With new theoretical advances in detector technology, in the coming year several experiments are poised to identify \cns~at reactors. The CONUS experiment is based at the $\sim 3.9$ Brokdorf reactor, and has released preliminary results and background measurements. The CONNIE experiment is using high-resistivity CCD detectors deployed at the Angra dos Reis nuclear power plant; their present upper limits are within approximately an order of magnitude of the predicted SM cross section at these energies~\cite{Aguilar-Arevalo:2016khx}. The MINER experiment is using a $\sim 1$ MW reactor source with a movable reactor core, and has published background predictions for their reactor set-up~\citep{Agnolet:2016zir}. The MINER set-up allows for detectors within close proximity to the core, $\sim 2-5$ meters, and for the reduction of systematics due to the unknown normalization of the reactor flux. Additional reactor experiments with unique sensitivies include RICOCHET~\cite{Leder:2017lva}, $\nu$-cleus~\cite{Strauss:2017cuu}, and TEXONO~\cite{Soma:2014zgm}.

\par Detecting \cns~from reactors will provide important constraints on beyond the SM physics that can be accessed through the low-energy neutrino sector. A future detection will be able to constrain NSI, and eventually place the most stringent bounds on the neutrino magnetic moment~\cite{Vogel:1989iv,Dutta:2015vwa}, as well as allow for a low energy determination of the Weinberg angle~\cite{Lindner:2016wff}. The detection will be complementary to the COHERENT detection in constraining NSI~\cite{ Coloma:2017ncl,Liao:2017uzy, Dent:2017mpr, AristizabalSierra:2018eqm, Denton:2018xmq}. This is because of the difference in the neutrino energies produced, as well as their flavor compositions. 

\section{Searches for new physics}
\par The new physics searches discussed up to this point have focused on NSI. However, dark matter experiments, COHERENT, and reactor experiments also provide a portal to beyond the SM ideas with new gauge and scalar interactions which may involve hidden sector particles. The new mediators probed by these experiments can range from a very low-mass scale ($\sim$ MeV) to a very high-mass scale $(\sim$ TeV). For comparison, the Large Hadron Collider (LHC) is most suitable for probing  mediators with mass $\gtrsim$ GeV.  

\par There are many well-motivated ideas for beyond the SM physics with a mediator mass in the range $\sim$  MeV to GeV. Dark matter candidates are viable in this mass range, and the mass of such a dark matter candidate may be associated to a sub-GeV hidden sector, or associated with explanations of the mass hierarchies of SM fermions~\cite{jk}. While this mass regime may be out of reach for experiments like the LHC and LEP, it is very suitable for the dark matter, COHERENT, and reactor-based experiments to probe. 

\par In this section we highlight and discuss three well-motivated scenarios for new physics which have particles in the MeV-GeV range: (i) kinetic mixing, (ii) hidden sectors, and  (iii) scenarios with a $L_\mu-L_\tau$ symmetry. We highlight the role of~\cns~and low energy neutrino experiments in probing these models. In addition, CE$\nu$NS experiments can also probe the parameter space of a fourth neutrino with mass $\Delta m^2\sim 1 \, \textrm{eV}^2$, which has been hinted at by several neutrino experiments but whose existence to this stage has proven inconclusive. 

\subsection{Kinetic mixing}
\par A simple way  to extend the SM is to add an extra $U(1)$ gauge group, which can arise in the context of Grand Unified Theories~\cite{Langacker:1980js}, string theory~\cite{Faraggi:1990ita}, as well as in hidden sectors~\cite{Mizukoshi:2010ky} that emerge in various phenomenological studies.  The new gauge group  with a gauge field $Z'$ and field strength $F'_{\mu\nu}$ mixes with the SM  via a  kinetic mixing of the form $\epsilon F'_{\mu\nu}F^{\mu\nu}$, where $F_{\mu\nu}$ denotes a SM field strength and $\epsilon$ parameterizes the strength of the mixing. The CE$\nu$NS process is particularly well suited for probing a light $U(1)$ gauge boson through mixing effects, where the couplings to the SM fermions are generated through kinetic mixings associated with the hypercharge boson and/or  mass mixings among $Z$ and $Z^\prime$ arising due to an extended Higgs sector (or via the Stueckelberg mechanism).

\begin{figure}[h]
\includegraphics[width=4.0in]{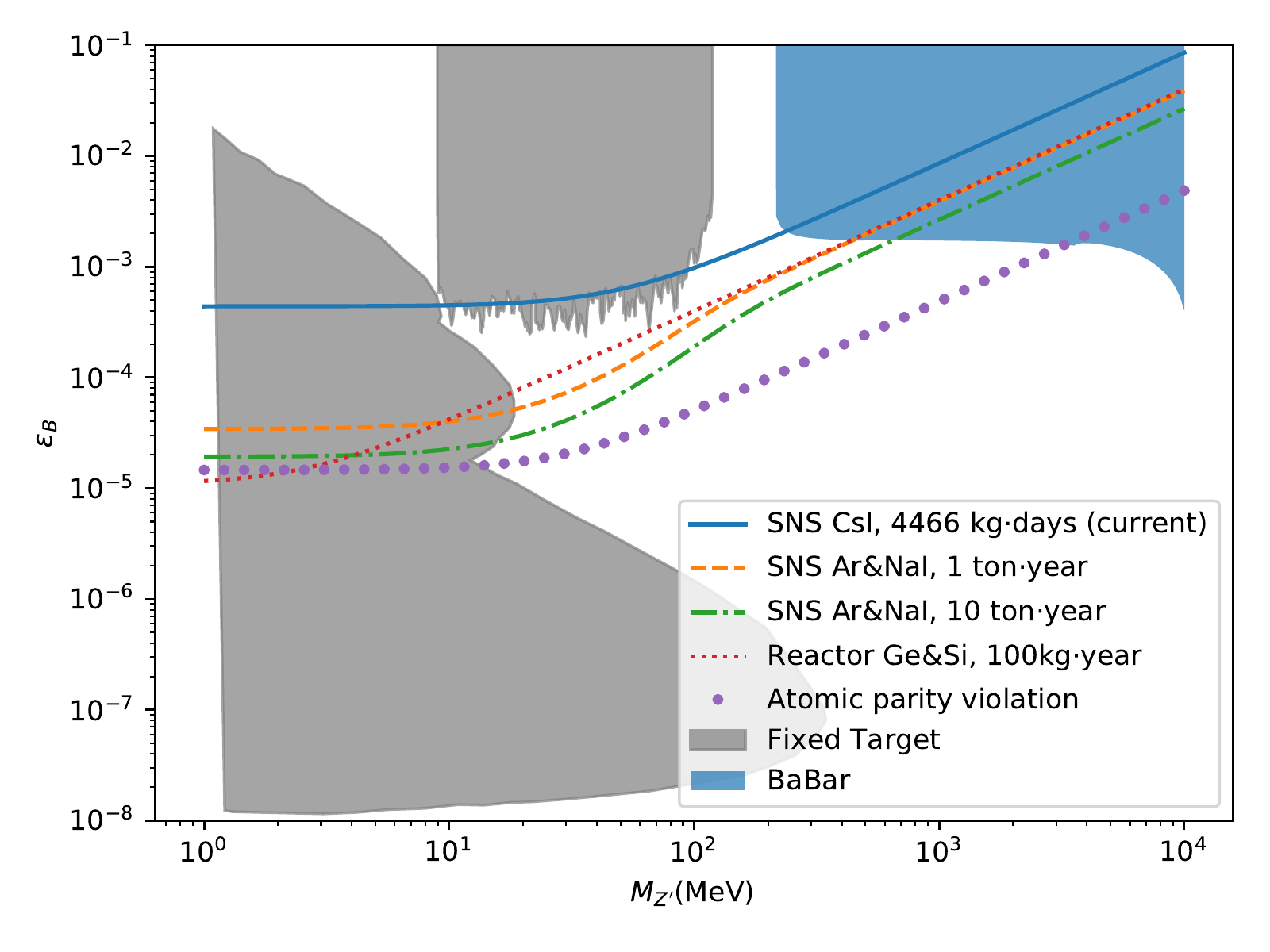}
\caption{The current and future bounds on the mixing $\epsilon_B$ in the dark hypercharge case are plotted as a function of the $Z'$ mass $M_{Z^\prime}$. The solid blue curve is the current COHERENT limit, the orange dashed and green dot-dashed curves are derived future projections for COHERENT for different exposures, the red dotted curved is the future projection for a reactor experiment, the purple large-dotted curve is from atomic parity violation, the grey regions are from the NA48/2, E774, E141, and E137 fixed target experiments. The blue shaded region is disallowed by the BaBar results. Figure reproduced from Ref.~\cite{Abdullah:2018ykz}.}
\label{fig:darkb}
\end{figure}

Figure~\ref{fig:darkb} shows the COHERENT and reactor reach on $\epsilon_B$ (associated with hypercharge-type dark boson) as a function of mediator mass $M_{Z'}$. The reach from COHERENT and reactor data is compared to limits arising from fixed target, atomic parity violation (APV) experiments, and the BaBar results. The region where the curves plateau is allowed by the fixed target experiments. Figure~\ref{fig:darkb} shows that current and projected limits from~\cns~measurements provide stringent constraints ($10^{-5} < \epsilon < 10^{-2}$) in the mass range $1\,\text{MeV}\, \lesssim m_{Z^\prime} \lesssim 10$ GeV, almost as strong as the existing limits from atomic parity violation. The BaBar results constrain $m_{Z^\prime}\sim 10$ GeV. Below about 10 MeV, the future COHERENT constraints are comparable to those arising from APV, and reactor experiments are projected to provide stronger limits due to the low energies of reactor neutrinos.

\begin{figure}[h]
\includegraphics[width=4.0in]{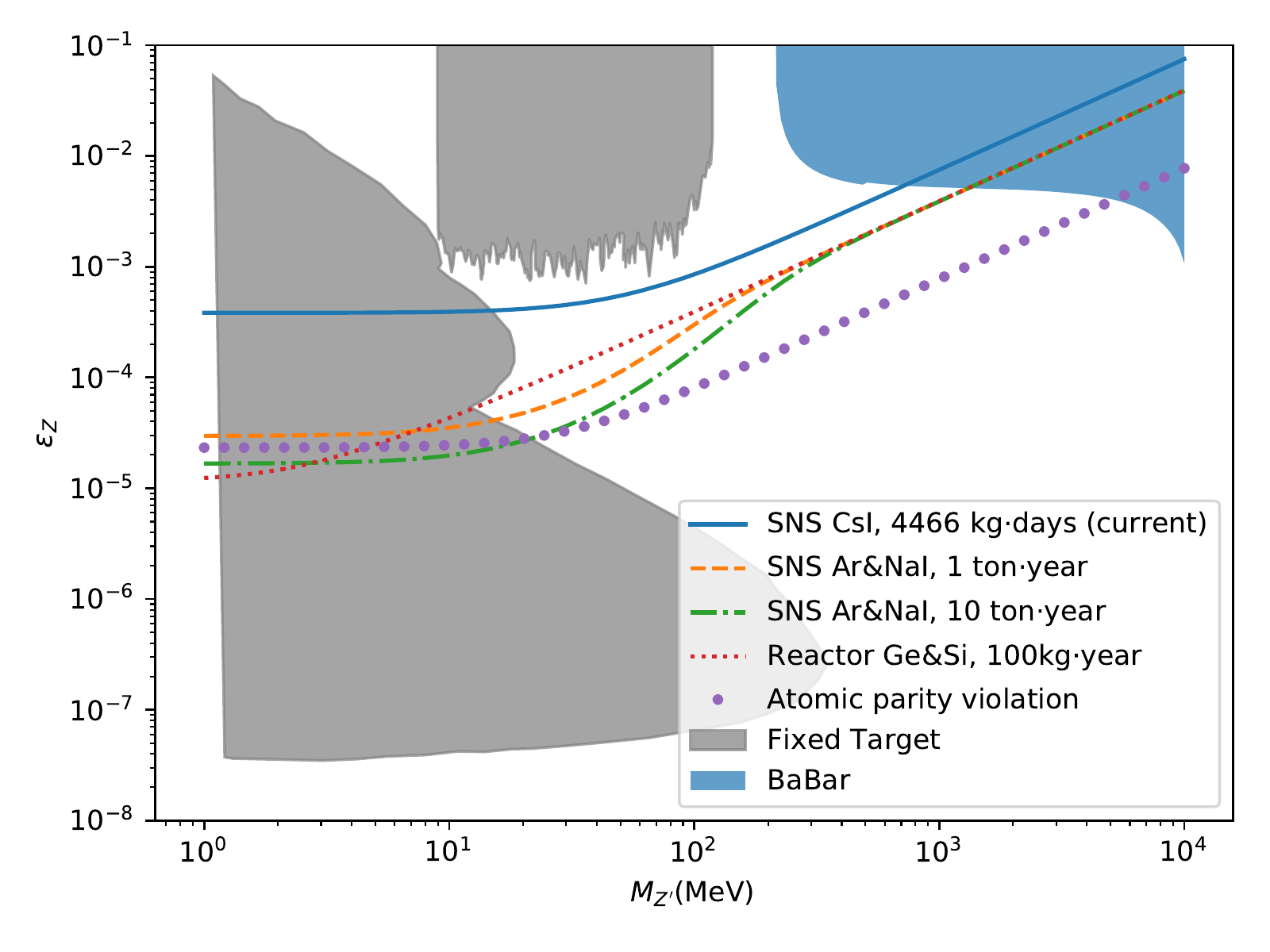}%
\caption{The current and future bounds on the mixing $\epsilon_Z$ in the dark $Z$ case are plotted as a function of the $Z'$ mass $M_{Z^\prime}$. The solid blue curve is the current COHERENT limit, the orange dashed and green dot-dashed curves are derived future projections for COHERENT for different luminosities, the red dotted curved is the future projection for a reactor experiment, the purple large-dotted curve is from atomic parity violation, and the grey regions are from the NA48/2, E774, E141, and E137 fixed target experiments. The blue shaded region is ruled out by the BaBar results. Figure reproduced from Ref.~\cite{Abdullah:2018ykz}.}
\label{fig:darkz}
\end{figure}

Figure~\ref{fig:darkz} shows the same constraints applied to $\epsilon_Z$ for the case of a dark $Z$ boson. The constraints are similar to the dark hypercharge case with two main differences. First, the window where the~\cns~ constraints start competing with APV is outside the bounds of the fixed target experiments. Second, for high values of $m_{Z'}$ the~\cns~limits become independent of the exposure and detector material. This effect is due to the nature of the $Z'$ coupling as well as the large exposure compared to the assumed systematic uncertainty. Note that bounds on dark $Z$ bosons may also be obtained from low-energy neutrino electron scattering~\cite{Lindner:2018kjo}. 

\subsection{Hidden sectors}
\par The interaction of quarks and leptons with a $Z^{\prime}$ can have an additional momentum transfer ($q^2$) dependence if these fermions couple to the $Z^{\prime}$ via a loop of hidden (dark matter) sector particles~\cite{Goodman:2010ku,Bai:2010hh,Fan:2010gt,Datta:2013kja,Elor:2018xku}. This new $q^2$-dependent term can be represented  by a form factor    $F(q^2)\sim q^2/\Lambda^2$, where $\Lambda$ is the scale in the dark matter (hidden) sector associated with the mass of the mediator particle that generates the quark-dark matter interactions, $\bar{q}q\bar{\chi}\chi$. As long as $\Lambda$ is greater than the maximum allowed momentum transfer for the scattering experiments, $F(q^2)\sim q^2/\Lambda^2$ appears in the scattering amplitude~\cite{Datta:2018xty}. 
 
\par
\begin{figure}
\centering
\begin{minipage}{.5\textwidth}
  \centering
  \includegraphics[width=2.9in]{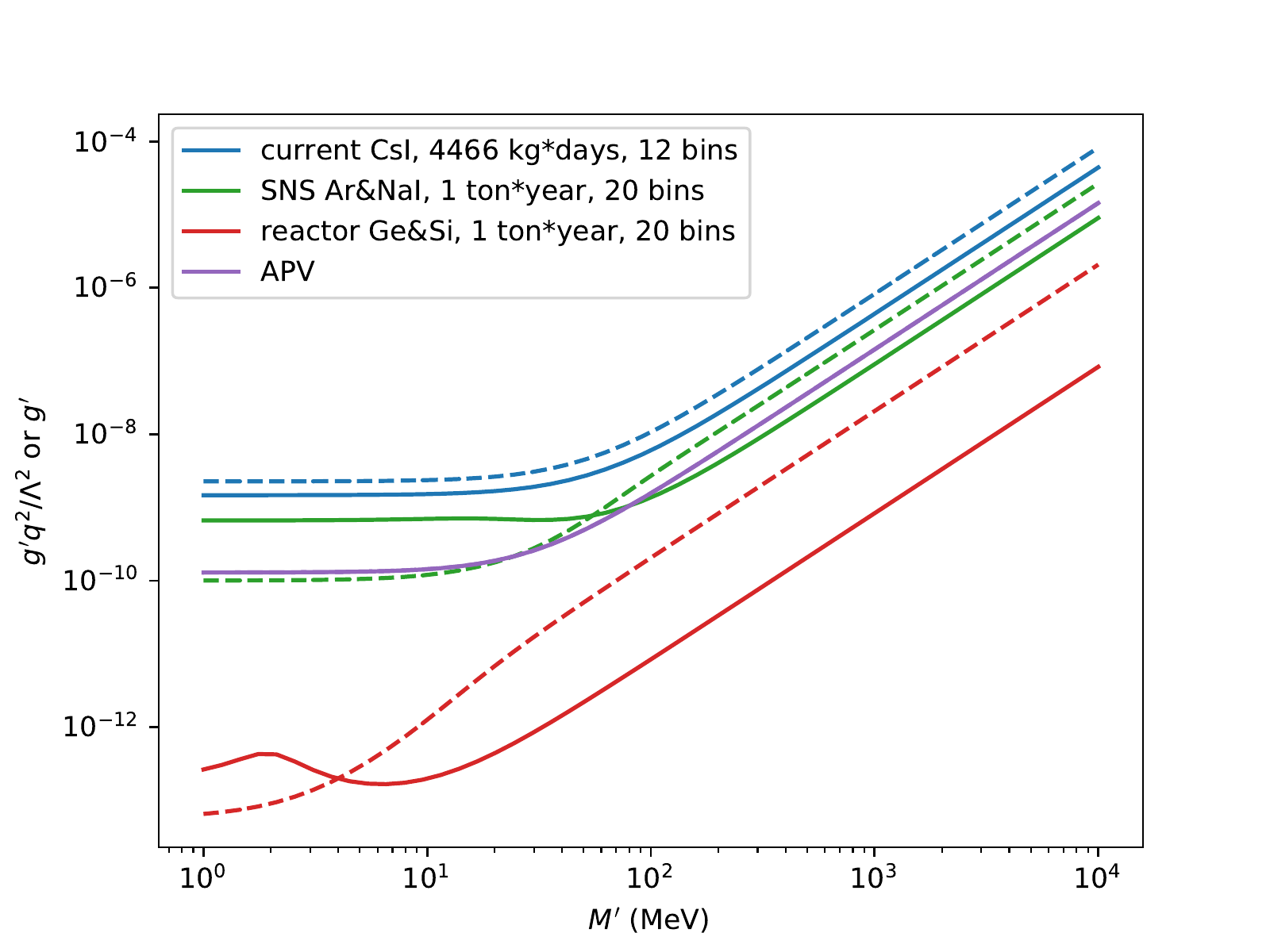}
\end{minipage}%
\begin{minipage}{.5\textwidth}
  \centering
  \includegraphics[width=2.9in]{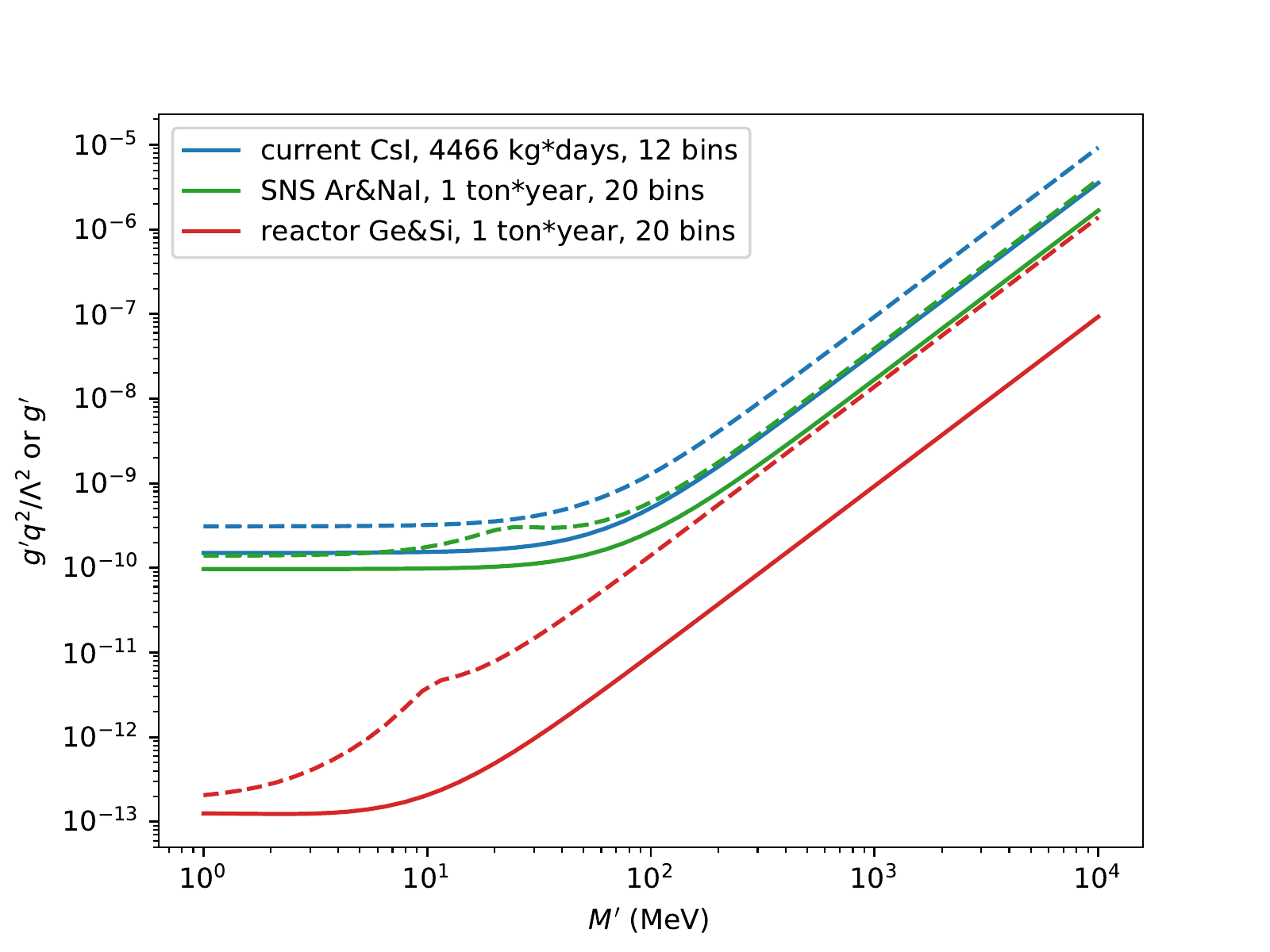}
\end{minipage}
\caption{Left: Current and projected 2$\sigma$ bounds on a vector (left) and scalar (right)  mediator with $F\left(q^2\right)\sim q^2$ as a function of the mediator mass. Dashed lines show the limits without a form factor. Here $q_0=50$~MeV for COHERENT, and $q_0=30$~MeV for reactor experiments. Figure reproduced from Ref.~\cite{Datta:2018xty}. }
\label{fig:scalarvector}
\end{figure}

 Figure~\ref{fig:scalarvector} shows the COHERENT and reactor constraints on the ratio,
\begin{equation}
    \frac{g^\prime q^2}{\Lambda^2}=\frac{((g_L+g_R) g_\nu) q_0^2}{2\Lambda^2}
    \label{eq:ratio}
\end{equation}
The constraints are shown at $2\sigma$ for a vector or scalar mediator, respectively, as a function of the mediator mass. The ratio shown in Equation~\ref{eq:ratio} represents the coupling strength between quarks and neutrinos as a function of energy which reduces to $g^\prime$ if there is no form factor for the coupling. A typical momentum transfer for these experiments, e.g. $q_0=50$ MeV and 30 MeV, are used for COHERENT and reactor experiments, respectively. For comparison the limits for the case without a form factor are shown as dashed lines. The quarks may contain direct couplings to the $Z$ and the couplings may emerge via dark matter loops in a given model, in which case the solid and dashed lines must be combined to obtain constraints on the couplings.  The bump in the low mass region for future COHERENT and reactor experiments is due to a combination of the form factor and the mediator propagator which yields $\frac{q^2}{q^2+m^{\prime 2}}\sim1$ and that the mediator-induced spectral distortion is suppressed. For the case without the form factor, the shape distortion persists for low masses, which makes the limits stronger compared to the $F\left(q^2\right)\sim q^2$ case. Note that  direct detection constraints are nonexistent for a sub-GeV DM mass and collider bounds are nonexistent for a GeV mediator which allows a lot of the parameter space to be unconstrained for $g\leq 1$. It is also interesting to note that the APV constraint is not very stringent for $F(q^2)\sim q^2$ case. 

\subsection{$L_\mu-L_\tau$}
\par Another simple and interesting extension of the SM invokes a non-universal $U(1)$ gauge symmetry associated with $U(1)_{L_\mu-L_\tau}$. This symmetry has been discussed in various contexts, especially those with the flavor structures of neutrinos \cite{He:1991qd,He:1990pn}, lepton flavor-violating Higgs decays \cite{Heeck:2014qea}, dark matter, and the recently reported flavor non-universality in B decays \cite{Altmannshofer:2016jzy}. This symmetry leads to a Lagrangian of the form:
\begin{equation}
\mathcal{L}_{int} \supset g_{Z^\prime}Q_{\alpha\beta}(\bar{l}_\alpha\gamma^\mu l_\beta+\bar{\nu}_{L\alpha}\gamma^\mu \nu_{L\beta})Z^\prime_{\mu},
\end{equation}
where, as before, $Z^\prime$ is the new gauge boson, $g_{Z^\prime}$ is the new gauge coupling, and $Q_{\alpha\beta}={\rm{diag}}(0,1,-1)$ gives the $U(1)_{L_\mu-L_\tau}$ charges. It is possible to extend this symmetry to the quark sector as well.  

\par Muon and tau loops at low energy generate kinetic mixing between the SM photon and $Z'$ of strength $\epsilon\propto (8e g_{Z^\prime})/(48 \pi^2)\text{log}(m_\tau/m_\mu)$ \cite{Kamada:2015era,Araki:2017wyg} (the $\mu$ and $\tau$ leptons can be replaced by  quarks if the symmetry is also extended to the  quark sector). Since this is generated at low energy, the diagonalization is done after electroweak symmetry is broken which produces a $Z'$ coupling to the first generation quarks equal to $\epsilon\,Q$. This mixing is suppressed by a loop factor, which is compensated for by the direct coupling to neutrinos in $\nu_\mu$ scattering experiments.

\begin{figure}[htbp]
\includegraphics[width=4.8in]{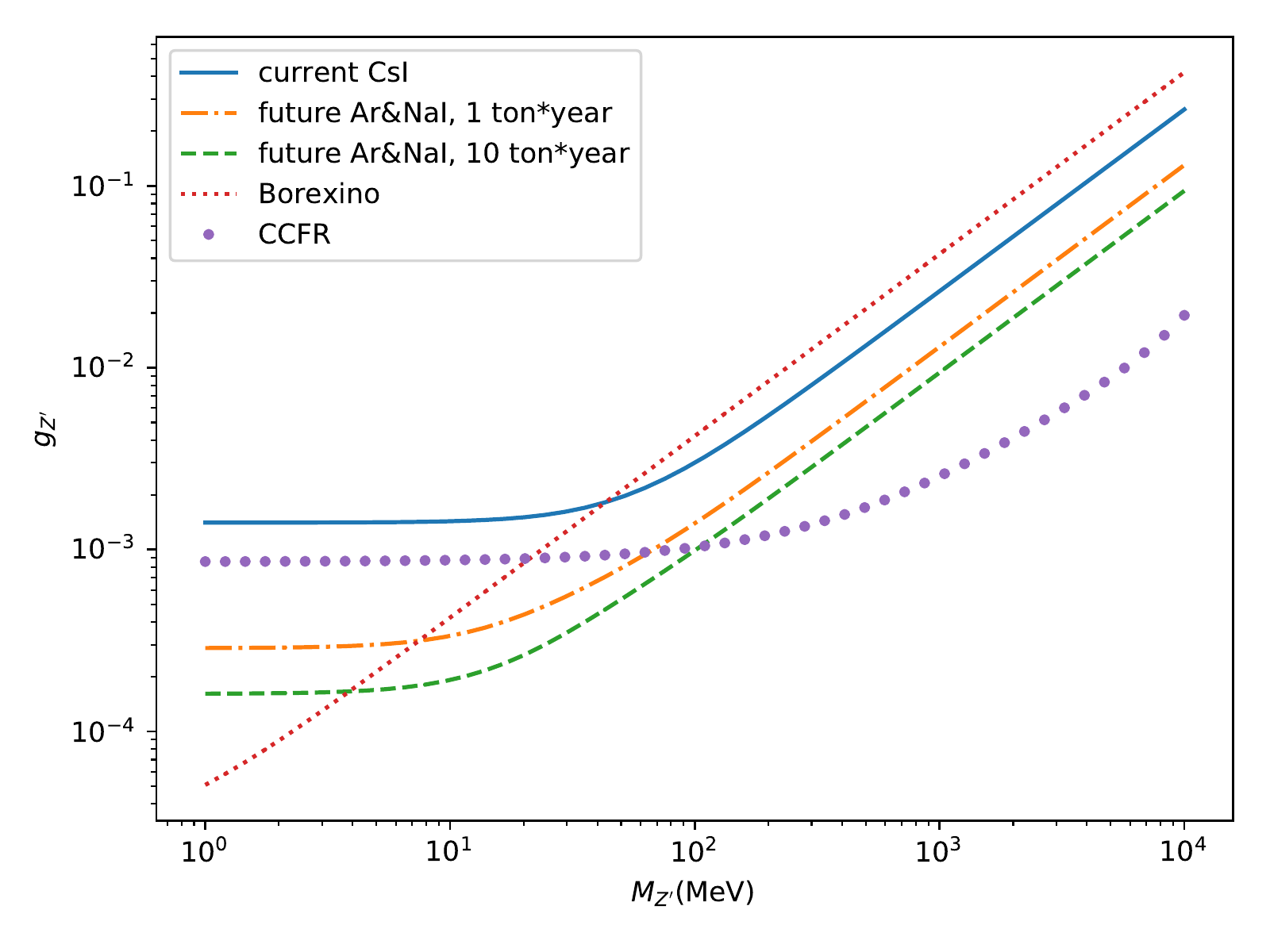}
\caption{Limits on $g_{Z'}$ as a function of $M_{Z^\prime}$ for $L_\mu$-$L_\tau$ models. The solid blue curve is the current COHERENT limit, the orange dot-dashed and green dashed are derived future projections for COHERENT for different detector exposures, the red dotted curve is from the Borexino measurement of solar neutrinos, and the purple large-dotted curve is from the CCFR measurement of neutrino trident production. Figure reproduced from Ref.~\cite{Abdullah:2018ykz}.}
\label{fig:lmultau}
\end{figure}

Figure~\ref{fig:lmultau} shows the limit on $g_{Z'}$ as a function of $M_{Z^\prime}$ for  $L_\mu$-$L_\tau$ models using current and projected COHERENT results, contrasting with the limits arising  from Borexino and CCFR. This shows that all these experiments are complementary, and in the mass window $4\,\text{MeV}\, \lesssim m_{Z^\prime} \lesssim 100$ MeV the future COHERENT projections provide the strongest limits ($g_Z'\, \leq \, 1-9 \times 10^{-4}$). Note that the reactor, fixed target, BaBar, and APV experiments have weaker limits in this scenario since they require electron flavor couplings. At the lowest $M_{Z^\prime}$  shown, the most stringent bounds come from the recent Borexino data~\cite{Agostini:2017ixy}. 

\subsection{Sterile neutrinos}
\par As discussed above several short baseline neutrino experiments hint at the presence of additional neutrinos beyond the three active components from the SM. Solar and reactor experiments have found indications of a deficit of electron neutrinos~\cite{Giunti:2006bj,Abdurashitov:2009tn,Kaether:2010ag,Giunti:2010zu}. Independently, very short baseline neutrino experiments with distances of $< 100$ m find evidence for a deficit of electron anti-neutrinos~\cite{Dentler:2018sju}. In addition the LSND and MiniBooNE~\cite{Aguilar:2001ty,Aguilar-Arevalo:2018gpe}
results can also be explained by the presence of sterile neutrinos.

\par At this stage it is simplest to interpret the data in terms of oscillation between an active neutrino and a single sterile component. The probability $P_{(\alpha\to\beta)}$ for oscillation between the two decoupled
neutrino flavors $(\alpha,\beta)$ is
\begin{equation}
P_{(\alpha\to\beta)} = \sin^2 \big[ 2 \theta \big] \times \sin^2 \big[ \frac{ \Delta m^2 L }{4 E_\nu }
\big] \,, 
\label{eq:oscprob}
\end{equation}
where $\theta$ is the mixing angle and the mass-squared difference $\delta m^2 \sim 1 \, \textrm{eV}^2$. For reactor experiments, the typical baseline is $\sim 1-20$ m, and for COHERENT the baseline is $\sim 20-30$ m for its current operating phase. This implies that the $L/E$ for these experiments are similar to that of MiniBooNE and LSND, so that they can probe the same sterile neutrino parameter space. 


The sensitivity reach of the COHERENT experiment for sterile neutrinos has been discussed in Ref.~\cite{Kosmas:2017zbh}. The MINER experiment has a configuration which makes it uniquely sensitive to sterile neutrinos; the projected reach of this experiment has been studied in Ref.~\cite{Dutta:2015nlo}. In particular, the MINER experiment, which utilizes detectors $\lesssim$ a few m from the reactor core, is complementary to the reach of LSND, Mini-Boone, and COHERENT in its constraint on the mass splitting $\Delta m_{14}^2$ $\sin^22\theta_{14}$. Future data from MINER should rival the reach of the on-going PROSPECT experiment~\cite{Ashenfelter:2018zdm}. 




\section{Summary}
\par In the near future, direct dark matter detection experiments will operate in a different phase. In the presence of astrophysical neutrinos, they will no longer be free from astrophysical backgrounds. In order to extract a possible dark matter signal, these backgrounds must be precisely measured. As we have discussed in this article, a measurement of this neutrino signal will be an interesting and unique scientific achievement in itself, opening up a new direction for new physics searches, and for extracting information from the neutrino sources. 

\par It is clear that whether a detection of astrophysical neutrinos or dark matter comes first, next generation detectors will be inherently multipurpose in their scientific scope. In addition to the dark matter and neutrino signals that have been discussed in this article, there are possible connections with other experimental endeavors. For the particular case of xenon detector, $^{136}$Xe could be depleted to aid in searches for dark matter and astrophysical neutrinos. This would benefit experiments which aim to observe neutrinoless double beta decay, which enrich the $^{136}$Xe of their target mass, and thus produce significant quantities of depleted xenon as a by-product. A future combined dark matter, neutrino, and neutrinoless double-beta decay observatory may be feasible if it can be designed to access a range of energies $\sim$ keV-MeV. 

\par Future dark matter detectors will be complementary to terrestrial experiments that are sensitive to $\sim$ MeV neutrinos. The recent detection of \cns~by the COHERENT experiment has provided novel bounds on beyond the SM physics, and motivates new theoretical developments in nuclear physics. Many interesting new physics scenarios, such as kinetic mixing, hidden sector models, flavor models, sterile neutrinos may now be investigated from a different perspective. New physics models with low mass mediators are especially interesting for~ \cns~, since they are not within the sensitivity reach of high energy colliders. Combining with an expected soon detection of~\cns~from reactor experiments, and employing at least two distinct target materials, can break a degeneracies that are not possible with a single type of experiment~\cite{Dent:2017mpr}. Due to the different sources of neutrinos that they are sensitive to, dark matter detection experiments are poised to improve even further on the sensitivity of terrestrial experiments. 

\par Even just focusing on the COHERENT experiment, extraction of information from this data is just beginning, with many more developments expected in the coming years. In the COHERENT experiment, muon type neutrinos and antineutrinos emerge from prompt and delayed decays of pion and muons, whereas the electron type neutrino arises from the  delayed decays of muons. Combining this timing information with the detected energy spectrum, it will be possible to distinguish electron type neutrino interactions from the muon type interactions when both types are present. This will lead to even stronger constraints on beyond the SM physics scenarios. 


\par In opening up a new experimental realm of low energy neutrino physics, it is interesting to consider what physics topics lie beyond those that we have discussed. For example, in our discussion of solar neutrinos, we have focused on the $\sim$ MeV neutrinos that are produced as by-products of nuclear burning in the center of the Sun. Even further down the road, we may ultimately consider thermally-produced neutrinos, which directly measure the temperature of the solar interior~\cite{Haxton:2000xb,Vitagliano:2017odj}. At the moment, the characteristic $\sim$ keV energies of these neutrinos is too far below the detection thresholds of modern experiments. Future detectors sensitive to these low energy recoils may also be able to probe the coherent scattering of neutrinos on atomic electrons~\cite{Sehgal:1986gn,Akhmedov:2018wlf}. Also at lower event rates than solar neutrinos, but in an energy realm that may be accessible to future experiments, geo-neutrinos~\cite{Gelmini:2018gqa} may provide a new window into the interior of the Earth. 

\section*{ACKNOWLEDGMENTS}
BD and LES acknowledge support from DOE Grant de-sc0010813. We would like to thank  M. Abdullah, J. Billard, A. Datta, J. Dent, E. Figueroa-Feliciano, G. Kane, R. Lang, S. Liao, D. Marfatia, J. Newstead and J. Walker for various collaborative works cited in this article. 

\bibliographystyle{ar-style5}

\end{document}